# An Approach to Secure Highly Confidential Documents of any Size in the Corporate or Institutes having Unsecured Networks

Samir B. Patel[1], Shrikant N. Pradhan[2]

Computer Science and Engineering Department, Institute of Technology, Nirma University, SG-Highway, Ahmedabad, Gujarat, India.
[1]samir.patel@nirmauni.ac.in, [2]snpradhan@nirmauni.ac.in

*Abstract*- **With the tremendous amount of computing because of the wide usage of internet it is observed that some user(s) are not able to manage their desktop with antivirus software properly installed. It is happening few times, that we allow our friends, students and colleagues to sit on our networked PC. Sometimes the user is unaware of the situation that there workstations are unsecured and so some one else could also be monitoring your flow of information and your most important data could go haywire, resulting into leakage of most confidential data to unwanted or malicious user(s). Example of some such documents could be question papers designed by the faculty member by various universities. Now a day most of the universities are having the biggest threat about the question papers and many other confidential documents designed by their faculty members. We in this paper present the solution to over come such a situation using the concept of Steganography. Steganography is a technique through which one can hide information into some cover object. This technique, if used, in positive direction could be of great help to solve such a problem and even other.**

## I. INTRODUCTION

The main objective of steganography is to avoid drawing suspicion to the transmission of a hidden message and make it seem "invisible" thus obscure the fact that a message is being sent altogether. For the purpose of steganography we can use an image, audio, video as the medium. We can embed secret information in the bit-planes of the image[1], in the samples of audio and frames of video.

The designed software provides one of the most secure approaches to communicate. In this design procedure the author has provided a facility with the intention that any data like images, text, any type of files and folders can be secretely communicated using cover objects like images, audio, video. The cover objects act as a vessel for the entire communication. Section 2 cover the basics of least significant bit position, section 3 and 4 introduces about the flow of the embedding and extraction modules respectively and section 5 provides the output screens generated by the system designed.

## II. BASICS OF LSB

In Least Significant Bit Steganography (LSB), any data is nothing more than strings and the string of bytes where in each byte is representing a different bit patterns. However, the last few bits in a byte do not hold as much significance as the most significant ones. This is to say that two bytes that only differ in the last few bits can represent two colors that are virtually impossible to differentiate by the human eye.

For example, 00100110 and 00100111 can be two different shades of red, but since it is only the last bit that differs between the two, it is impossible to see the color difference. LSB Steganography, then, alters these last few bits by hiding a message within them. The last bit of every byte is replaced with the corresponding bit.

The new image now contains the desired text without degrading the quality of the image since only the least significant bits were altered.

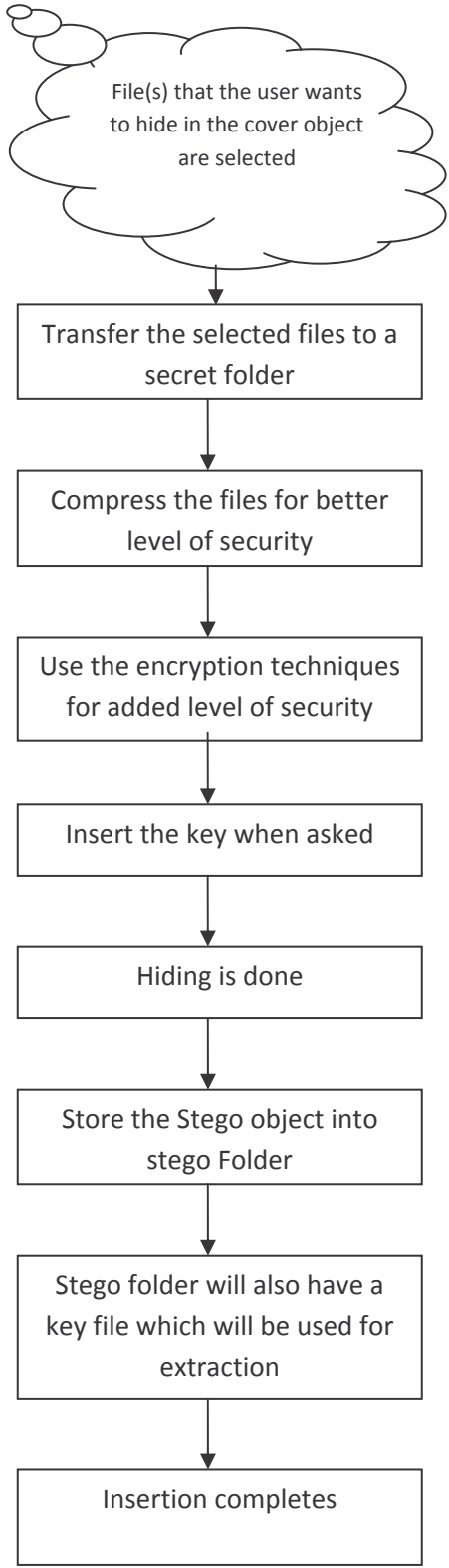

Figure 1. Complete flow for the Embedding process

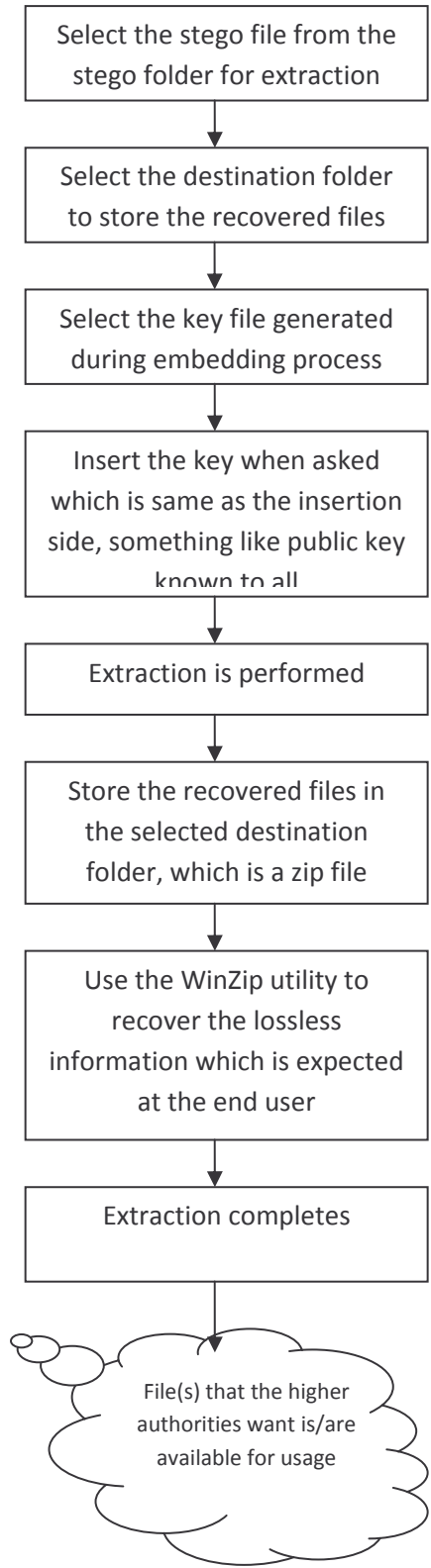

Figure 2. Complete flow for the Extraction process

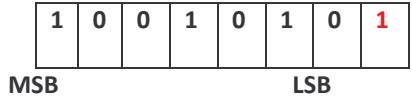

**MSB**              **LSB**

Usually a pixel size of 24-bits or 8-bits is used to store a digital image. The former one provides more space for information hiding; however, it can be quite large in size [1, 3, 4, 6].

The colored representations of the pixels are derived from three primary colors red, green and blue. 24-bit images use 3 bytes for each pixel, where each primary color is represented by 1 byte. Using 24-bit images each pixel can represent 16,777,216 color values.

We can use the lower two bits of these color channels to hide data, then the maximum color change in a pixel could be of 64-color values, but this causes small change that is undetectable for the human vision. This simple method is known as Least Significant Bit insertion .Using this method we can embed information with no visible degradation of the cover image, for more information in this regards [7].

### III. EMBEDDING PROCESS

The complete flow chart for the embedding is as shown in Figure 1. Any file or group of files including image, document, text, audio etc. could be embedded using the developed tool.

### IV. EXTRACTION PROCESS

The complete flow chart for the extraction i.e. decoding is shown in Figure 2. The screens indicating the complete flow is shown below as an experimental outcome with various steps mentioning what is done during each step.

Security involved in this approach is very high - because no information is traveling in raw form, all the information is encrypted and then converted into zip file, so even though the intruder finds the zip file, as the complete data is encrypted with the most secure algorithms and keys involved in it, it is not only difficult but impossible to get the actual data. And, one must not forget that the data is flowing within the cover object which is not showing any indication that something is hidden within it.

### V. EXPERIMENTAL RESULTS

Table1 and Table2 shows the sample file size and the amount of time involved during the embedding and extraction process. Looking into Table1 first entry of 843 Kile Byte file size the time required to do embedding in the cover object including compression and security algorithm execution is 25 Seconds. Like wise for the other files of different sizes the time taken is mentioned in the Table. Lesser the size less time the algorithms will take to embedded the whole file(s) in the cover object. Different file sizes are considered and shown below for more clarity with RSA and DES as algorithms considered for encryption and decryption of few files.

**Time Analysis:**

TABLE1. Encryption + Decryption (RSA)

| Size | Time required to be completed |
|------|-------------------------------|
| 843kb | 25s |
| 167kb | 4.5s |
| 3.5mb | 1min20sec |
| 100k | 2.85s |

TABLE2. Encryption + Decryption (DES)

| Size | Time required to be completed |
|------|-------------------------------|
| 843kb | 0.25s |
| 167kb | 0.005s |
| 16mb | 4s |

The steps listed at the end of the paper gives the complete understanding of the complete work.

## VI. CONCLUSION

There are many tools available for performing steganography work, but the work that we have carried out is unique in the way that, any file or group of files or even a complete folder containing any type of data of any size could be very easily embedded into any of the cover object including image, audio or video. Also encryption and decryption is fully exploited to undergo the implementation work for having maximum level of security build inside the package.

The package is tested successfully with all the cover objects and was found to be very user friendly to use for embeeding the documents into the multimedia object. We have successfully tested all the different types of documents, including images, videos, audio, plain text, document files with different extensions and were successful to hide and retrieve all the confidential information/documents using our package without any error.

Future work could include usage of the transformation based techniques like Discrete Cosine Transformation (DCT) and Discrete Wavelet Transformation (DWT) based implementations to get more robustness against the attack [8][9][10].

## ACKNOWLEDGEMENT

We would like to thank all the members of NIRMA UNIVERSITY for providing continuous support and inspiration.

We are thankful to Mr. Jaldeep Vasavda , Mr. Nirav Shah and Mr. Sandip Patel for providing the programming support when needed.

### AUTHORS PROFILE

**First Author:** Prof. Samir B. Patel is born in Ahmedabad, 26/07/1975 , He obtained his degree BE computer Engineering from LD college of Engineering, Ahmedabad, Gujarat, India in the year 1998, He did his M.E computer Engineering from SP University, Vallabh Vidhyanagar, Gujarat, India. His area of interest involved processing multimedia data, data Security, parallel computing. He is currently working at the post of Sr. Associate professor at CSE Department, Institute of Technology, Nirma University, Ahmedabad, Gujarat, India.

**Second Author:** Dr. Shrikant N. Pradhan is Ph.D. and he is having a vast experience of research. He was associated with Physical Research Laboratory (INDIA) for more than 25 years. He is currently Head of the M. Tech Programme at Institute of Technology, CSE Department, Nirma University., Ahmedabad, Gujarat, India. His area of interest involves signal processing, Embedded systems, Multimedia, Data Security and many more.

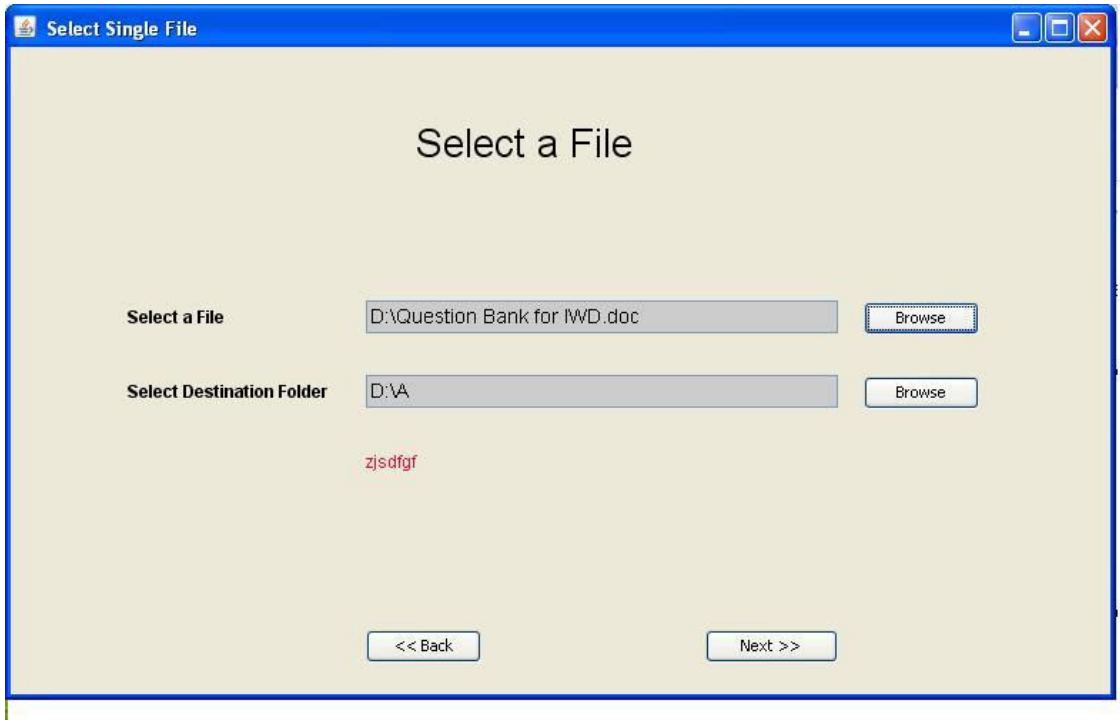

Figure 3. Step 1 selecting a file that the user wants to hide and also select the destination folder, here by selecting the destination folder, the files that the user wants to hide will move to this secret folder.

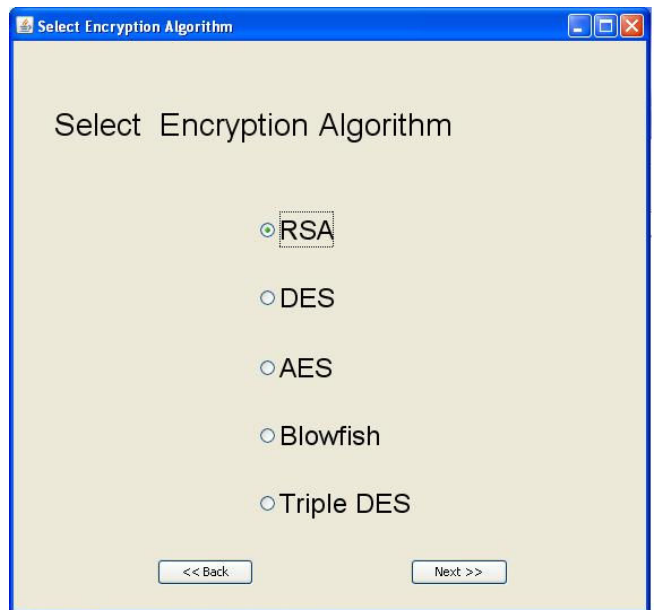

Figure 4. Step 2: Here the user has to select the encryption algorithm of choice and according the encryption will be carried out, to have one of the best levels of security this module is integrated.

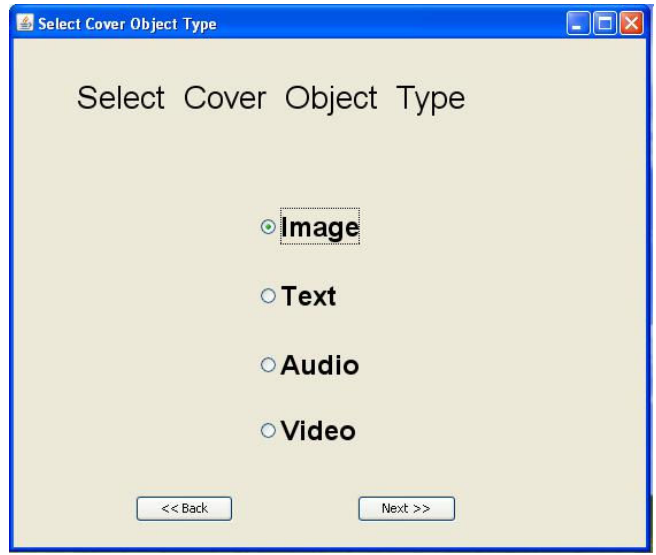

Figure 5. Step 3: Select the cover object of specific type for the type of data that the user wants to hide.

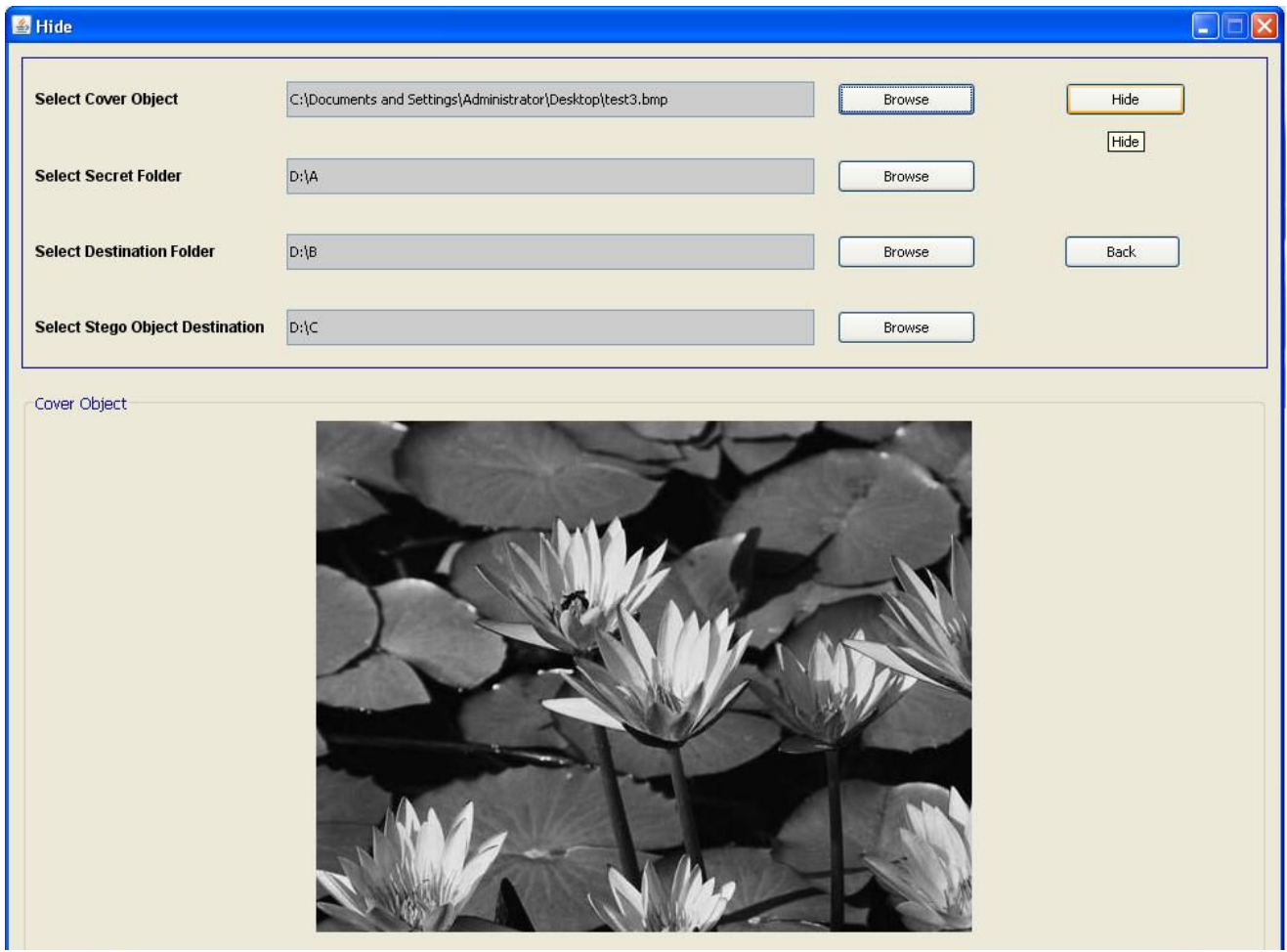

Figure 6. Step 4: Here the selection of the various files and folder is to be done so that the embedding can be carried out. A sample data is shown here for convenience. One must have the appropriate folder created beforehand so that those folders can be utilized for the actual need.

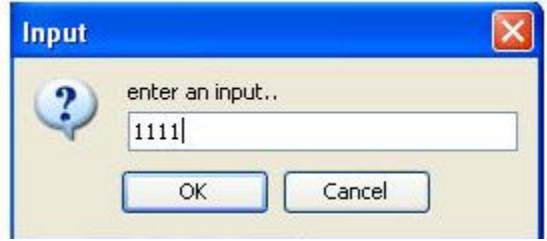

Figure 7. Step 5: Select the agreed upon key of your choice. This key number that is entered must be the same on the extraction side; otherwise the algorithm will not be able to extract the same information and will be a total chaos.

Step 6: Hiding done dialog box will appear.

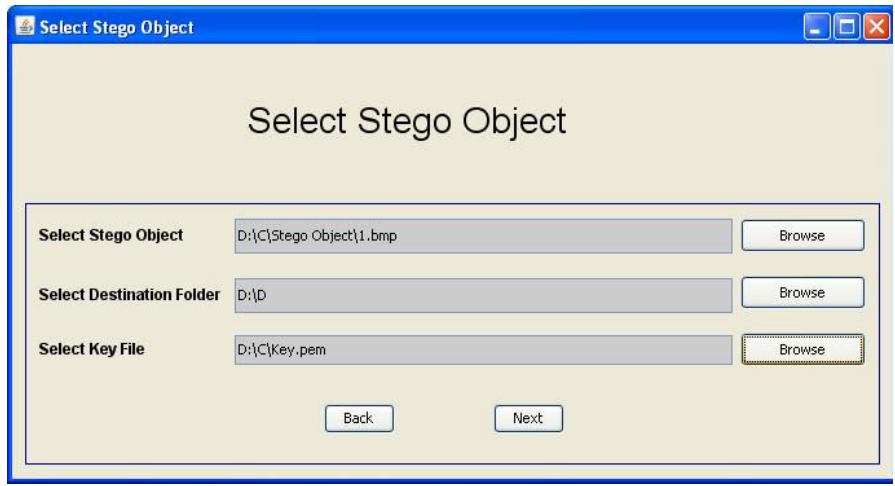

Figure 8. Step 7: This the decoder side i.e. extraction side where in the stego objects where ever it is stored must be provided to the system and also the user has to select the destination folder and the key file which is generated by the system. Without this key file the extraction module will not work. This means that even though the intruder is having the input number as in step5 but if it does not have a key then those numbers are of no use.

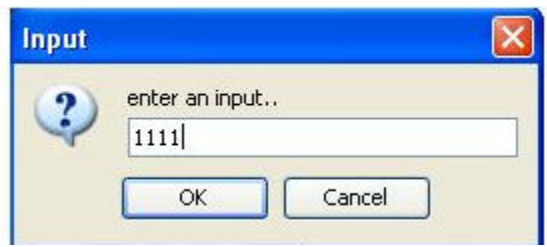

Figure 9. Step 8: The system will ask for the key number input which is the same as in step five and is only shared between the two ends.

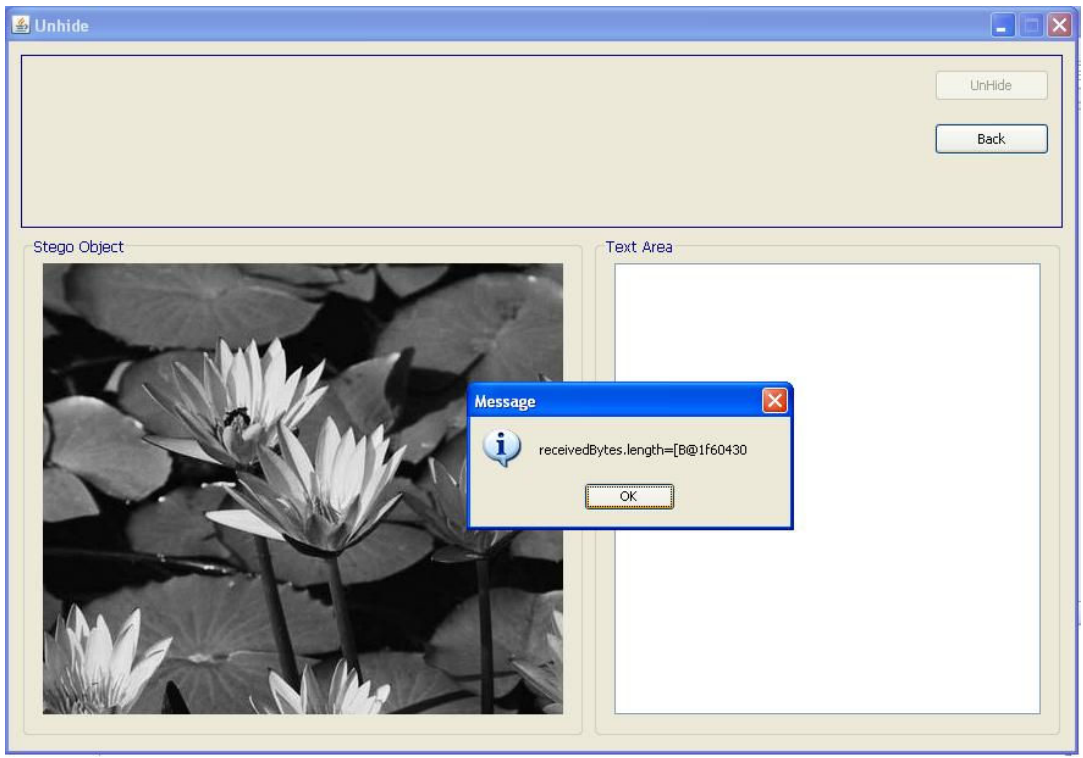

Figure 10. Step 9: Once you press the unhide button the total length of data received will appear in the dialog box.

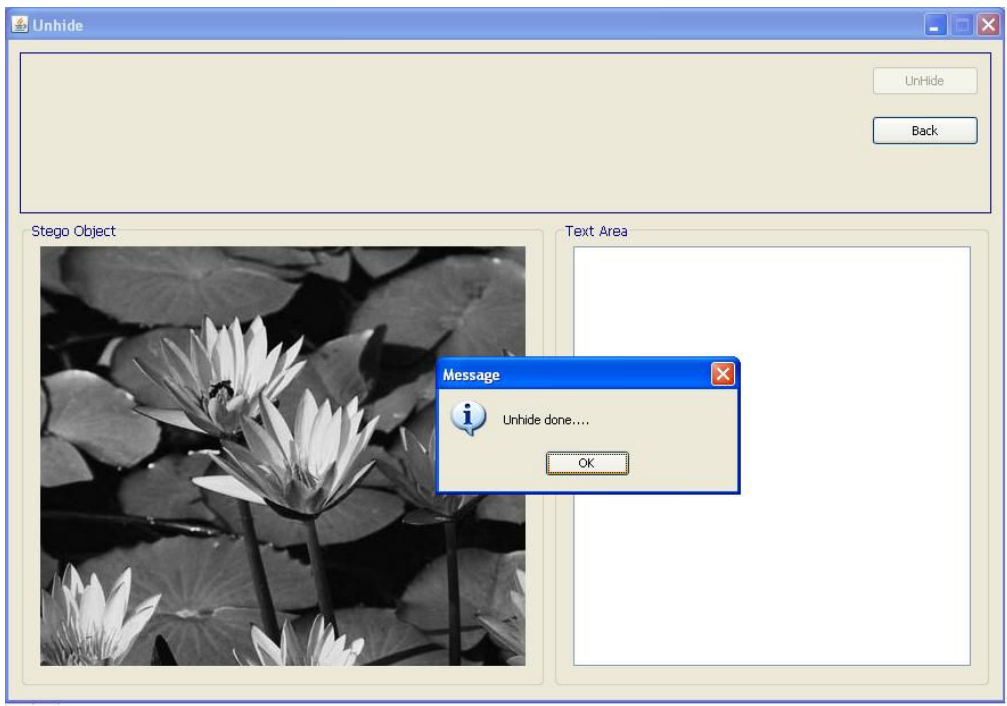

Figure 11. Step 10: Un-hide done dialog will appear on the screen and this is the last process.

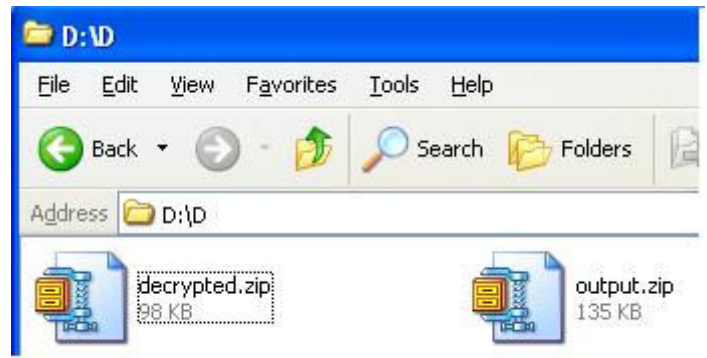

Figure 12. Step 11: Complete extraction is done as specified in the folder selection during the step7 and so finally the decrypted.zip file will be available which contains the actual document. This document could be your most secret question paper.

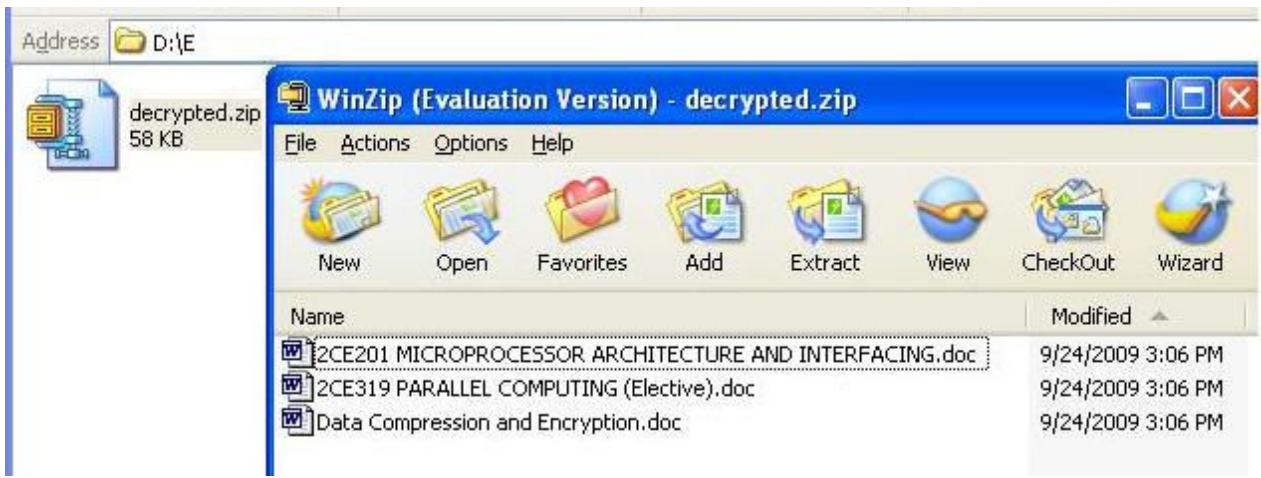

Figure 13. Step 12: Use the WinZip utility to uncover the decrypted.zip file which contains the necessary documents which the source have embedded in the cover object. Using the stegoed cover objects the recovery is done inside the selected destination folder where this decrypted.zip file is present, you have to get the required files from this by unzipping the file.